\documentstyle[12pt,preprint,prd,aps]{revtex}
\begin{document}

\draft
\begin{titlepage}
\preprint{\vbox{\hbox{IASSNS-AST 96/58}
\hbox{SISSA 166/96/EP}
\hbox{hep-ph xxxxxxx}
\hbox{November 1996} }}

\title{Constraints on Energy Independent
Solutions of the Solar Neutrino Problem}

\author{P.I. Krastev$^{a~*}$ and S.T. Petcov$^{b}$ \footnote{Also at:
Institute of Nuclear Research and Nuclear Energy, Bulgarian Academy of
Sciences, 1784 Sofia, Bulgaria.}}

\address{$^a$ School of Natural Sciences, Institute for Advanced
Study, \\ Princeton, NJ 08540 \\}

\address{$^b$ Scuola Internazionale Superiore di Studi Avanzati, and
\\ Istituto Nazionale di Fisica Nucleare, Sezione di Trieste, I-34013
Trieste, Italy \\}

\maketitle

\begin{abstract}
We analyze the latest published solar neutrino data assuming an
arbitrary neutrino oscillation/conversion mechanism suppresses the
electron neutrino flux from the Sun independent of energy. For
oscillations/transitions into active (sterile) neutrinos such
mechanisms are ruled out at 99.96 (99.9997) \% C.L. assuming the
standard solar model by Bahcall and Pinnsoneault '95 correctly
predicts all solar neutrino fluxes within their estimated
uncertainties. Even if one allows for $^8{\rm B}$ and $^7{\rm Be}$
solar neutrino fluxes that are vastly different from the ones in
contemporary standard solar models these mechanisms are strongly
disfavored by the data.
\end{abstract}
\end{titlepage}

\section{Introduction}
\label{Intro}

The present solar neutrino data \cite{CHLOR,KAM,GALLEX,SAGE}, are in
significant disagreement with the most elaborate contemporary standard
solar models (SSM) \cite{BP95,CDFLR,RVCD}. Theoretical uncertainties
and experimental errors cannot account for this discrepancy and the
data indicate that the solar $\nu_e$ flux is reduced with respect to
its predicted value in the SSM.  Assuming the electron neutrinos
($\nu_e$) produced in the Sun and later detected in the solar neutrino
detectors on Earth, travel unaltered between the points of production
and detection, and using the luminosity constraint one finds that the
inferred from the data reduction of the intermediate energy $^7{\rm
Be}$ neutrinos is much stronger than the reduction of both the
low-energy $pp$ neutrinos, the flux of which actually appears to be
increased by 10 \% at the best fit point, and the high-energy $^8{\rm
B}$ neutrinos, the flux of which turns out to be lower by a factor of
$\simeq 0.3$ \cite{indep}.

Presently the only known way to explain the discrepancy between the
predictions of the solar models and the experimental results is to
resort to neutrino flavor conversion mechanisms (MSW \cite{MSW},
vacuum oscillations \cite{VAC}, etc.)  which can provide an energy
dependent suppression of the known solar $\nu_e$ fluxes ($pp$, $^7{\rm
Be}$, $^8{\rm B}$, $pep$, $^{13}{\rm N}$ and $^{15}{\rm O}$).  An
important role in the explanation plays also the $\nu_\mu$ (and/or
$\nu_\tau$) neutral current contribution to the neutrino-electron
scattering rate, present in the case of solar $\nu_e$
oscillations/transitions into active neutrinos. This contribution
increases the ratio of the event rate in the Kamiokande detector to
the event rate expected from the SSM without oscillations, with
respect to the analogous ratio for the chlorine experiment, in
accordance with the data.

In a previous publication \cite{paperI} we have shown that neutrino
conversion mechanisms which predict an energy independent suppression
\footnote{Here and in what follows we always have in mind solar
neutrino conversion mechanisms which lead to an energy independent and
equal suppression of all components of the solar $\nu_e$ flux.  For
brevity we call such mechanisms ``energy independent mechanisms''. Let
us note that, for instance, the resonant solar neutrino transitions
due to the interplay of neutrino flavour changing and flavour diagonal
(but flavour dependent) neutral current interactions \cite{FCNC} are
in the case of massless neutrinos energy independent.  Nevertheless,
the corresponding transition probabilities can be different for the
$^{8}$B, $^{7}$Be, $pp$, etc.  neutrinos (see ref. \cite{FCNC} for
details).}  of the entire solar neutrino spectrum are ruled out by the
data, assuming the standard solar model \cite{BP92} correctly predicts
all neutrino fluxes. We also showed that even if the $^{8}$B neutrino
flux is reduced by one half and the $^{7}$Be neutrino flux is reduced
by 30 \% with respect to the corresponding theoretically predicted
fluxes in the SSM, these mechanisms still provide a poor fit to the
data. Examples of such mechanisms with energy independent suppression
are neutrino spin-precession \cite{VVO}, vacuum oscillations with an
oscillation length much shorter then the Sun - Earth distance,
adiabatic MSW transitions for all neutrino flux components, possible
only for large mixing angles and $10^{-7}~{\rm eV}^2 \leq \Delta m^2
\leq 10^{-5}~{\rm eV}^2$, etc.

Our previous analysis has been published as a part of a larger
investigation which was centered on somewhat different problems, used
older data and by now an outdated reference solar model. Here we
update our previous results by using the latest solar neutrino data
summarized in Table. \ref{SNData}. We extend our analysis by using
neutrino fluxes from three recent solar models \cite{BP95,CDFLR,RVCD}
which include helium and heavy elements diffusion, an effect that has
been shown lately to considerably improve the agreement with
helioseismological very precise measurements of low degree p-modes
\cite{helio,BPin}. We also use recently updated cross-sections for the
chlorine and gallium detectors \cite{BLABFN} which improve the
accuracy of the predicted event rates in these detectors.  We explain
in detail the assumptions and numerical procedures used in the
analysis. Our latest results confirm and strengthen the results and
the conclusions reached in our previous publication \cite{paperI}.

\section{Numerical Procedure}
\label{NumProc}
Given the published experimental results, a standard solar model which
predicts the neutrino fluxes, and a mechanism depleting the fluxes of
solar $\nu_e$ by a factor $P$ which is independent of energy, one can
compute $\chi^2$ including the theoretical uncertainties and
experimental errors as described, e.g., in \cite{fogli}. The
theoretical uncertainties are due to uncertainties in the input
parameters of the standard solar model as well as to uncertainties in
the detection cross-sections. The correlations between the solar
neutrino fluxes which arise because one and the same input parameter
can affect several neutrino fluxes also have to be taken into
account. In this calculation we have only one independent parameter,
$P$, which is supplied by the relevant mechanism proposed as a
solution of the solar neutrino problem. We describe the results of
this calculation in the next section.

We next relax the uncertainty intervals, estimated within the SSM, of
two of the most uncertain solar neutrino fluxes, $^8{\rm B}$ and
$^7{\rm Be}$.  For three representative values of the beryllium
neutrino flux, $\Phi_{Be}$, equal to 0.7, 1.0 and 1.3 times the flux
in the SSM we vary both $P$ and the total boron neutrino flux,
$\Phi_B$, treating them as free parameters \cite{paperI}. This
procedure was adopted in order to study how much model dependent are
the strong constraints, obtained with the first procedure described
above, on the energy independent neutrino oscillation/conversion
mechanisms. One of the largest uncertainties of the solar model
predictions is related to the fact that the cross-section for the
$^7{\rm Be}(p,\gamma)^8{\rm B}$ reaction has never been measured at
the typical energies at which this reaction is taking place in the
Sun. This is the primary source of the large existing uncertainty in
the predicted boron neutrino flux. In the model \cite{BP95} the
estimated one sigma relative uncertainty in the $^8{\rm B}$ flux is
$({}^{+0.14}_{-0.17})$ and the corresponding uncertainty in the
$^7{\rm Be}$ neutrino flux is $({}^{+0.06}_{-0.07})$. The relative
uncertainties in the SSM prediction of the CNO neutrino fluxes,
$^{13}{\rm N}$ and $^{15}{\rm O}$, are even larger than those of the
boron flux. However, because of their lower energies and relatively
low fluxes, the CNO neutrinos contribute much less to the event rates
in the chlorine and gallium neutrino experiments than the $^{8}$B
and/or $^{7}$Be neutrinos and thus are less important in the analyses
of the solar neutrino data. Therefore in our calculations we keep the
CNO neutrino fluxes equal to their SSM values.

The predicted event rates in the operating detectors can be expressed
as functions of the corresponding event rates as calculated from the
SSM, the constant suppression factor, $P$, and the assumed ratios,
$f_\alpha$, of each solar neutrino flux, $\Phi_\alpha$, to the same
flux in the SSM \cite{BP95}, $\Phi_{\alpha,SSM}$, $f_\alpha =
\Phi_\alpha/\Phi_{\alpha,SSM}$, $\alpha = pp$, $^7{\rm Be}$, $pep$,
$^8{\rm B}$, $^{13}{\rm N}$ and $^{15}{\rm O}$. The three equations
which give these event rates are:

\begin{mathletters}
\label{eq:all}
\begin{eqnarray}
Q_{Ga} & = & P\left(f_B Q_{Ga}^{B} + f_{Be}Q_{Ga}^{Be} +
f_{pp}Q_{Ga}^{pp} + f_{pep}Q_{Ga}^{pep} + Q_{Ga}(CNO)\right),
\label{eq:a} \\
Q_{Cl} & = & P\left( f_B Q_{Cl}^{B} + f_{Be}Q_{Cl}^{Be} +
f_{pep}Q_{Cl}^{pep} + Q_{Cl}(CNO)\right), \label{eq:b} \\
R_{\nu e}& = & f_B(0.845 P_B + 0.155), \label{eq:c} \\
or \nonumber \\
R_{\nu e}& = & f_B P_B. \label{eq:d}
\end{eqnarray}
\end{mathletters}
The quantities in the lefthand side of the first two equations are the
rates of $^{71}{\rm Ge}$ and $^{37}{\rm Ar}$ production in the gallium
and chlorine detectors and $R_{\nu e}$ in Eq. (\ref{eq:c}) and
Eq. (\ref{eq:d}) is the ratio (measured $^8{\rm B}$ flux)/(SSM $^8{\rm
B}$ flux) for the Kamiokande $(\nu_e e)$ scattering experiment.
Depending on the particular mechanism involved, $P$ can be the solar
$\nu_e$ survival probability for neutrino oscillations, spin-flavor
conversion, etc.  For neutrino oscillations/transitions into active
neutrinos ($\nu_\mu$ and/or $\nu_\tau$), or antineutrinos
($\bar\nu_\mu$ and/or $\bar\nu_\tau$), Eq. (\ref{eq:c}) is the
relevant one, whereas Eq. (\ref{eq:d}) has to be used for
oscillations/transitions into sterile neutrinos ($\nu_s$).
$Q^{\alpha}_i$ is the contribution of source $\alpha$ to the total
event rate in the i-th radiochemical detector ($i = Ga, Cl$) as
predicted in the SSM. The coefficients $f_\alpha$ are written
explicitly only for those fluxes for which we have assumed that they
might be different from the ones predicted in the SSM.

For $f_{Be} =0.7$ (1.3) we chose to satisfy the luminosity constraint
including small corrections for thermal motion of the particles in the
plasma and inequalities between the neutrino fluxes, which follow from
the sequencial process of nuclear fusion in the Sun \cite{BK}
\footnote{For earlier discussions of the luminosity constraint
see ref. \cite{lumconst}.}, by increasing
(decreasing) the $pp$ neutrino flux by $\simeq 2$ \%.  We always kept
the ratio of the $pep$ to $pp$ flux equal to the one in the SSM.
The prediction for this ratio in the different solar models is
remarkably stable: it doesn't vary by more than one percent. The
numerical coefficients in Eq. (\ref{eq:c}) have been computed using
the neutrino-electron scattering cross-sections with radiative
corrections \cite{sirlin}. The ratio (0.155) of the (integrated over
the $^8{\rm B}$ spectrum) scattering cross-sections $\sigma(\nu_\mu
(\nu_\tau) e)/\sigma(\nu_e e)$ has been computed using the published
trigger efficiency functions and energy resolution of the detector. We
have taken into account also the known evolution of the trigger in
Kamiokande \cite{KAM}. During the first 449 days the detectors has
been operated with a threshold of 9.3 MeV, later on the threshold has
been reduced to 7.5 MeV (794 days) and 7.0 MeV (836 days). A slight
change in the trigger efficiency function during the Kamiokande-II
stage due to increased light collection efficiency has been
neglected. This effect might change the time-averaged ratio at most by
a few percent, which will not change significantly any of our results.

Next we computed the $\chi^2$ defined as:

\begin{equation}
\chi^2 = \sum_{i=1}^{4} \frac{(Q_{exp}(i) -
Q_{th}(i))^2}{\sigma_{exp}^2(i) + \sigma^2_{th}}
\label{chi}
\end{equation}

\noindent The index $i$ denotes the four operating experiments. The
statistical and systematic errors from each experiment have been
combined quadratically to find the total experimental error,
$\sigma_{exp}(i)$.  The theoretical error is due to the uncertainties
in the solar $\nu_e$ detection cross-sections for the two
radiochemical experiments \cite{BLABFN} and has been calculated as
described in \cite{fogli}. The theoretical uncertainty for the
Kamiokande experiment has been assumed to be zero.  Here we neglected
the theoretical uncertainties related to the input parameters in the
solar model but instead varied two of the most uncertain fluxes,
$\Phi_B$ and $\Phi_{Be}$, within much broader intervals than their
estimated 3$\sigma$ uncertainty ranges in the SSM.  This approach
allows us to do the same analysis with three independently developed
standard solar models \cite{BP95,CDFLR,RVCD}.

\section{Results}
\label{results}
Assuming the standard solar model \cite{BP95} correctly predicts,
within the estimated theoretical uncertainties, the flux from each
solar neutrino source, the energy independent mechanisms provide a
very poor fit to the data. For oscillations/transitions into active
neutrinos the minimum $\chi^2$ is 13.0 (1 d.f.) which rules out this
scenario at 99.96 \% C.L. The constraint on oscillations/transitions
into sterile neutrinos are even stronger.  The minimum $\chi^2$ is
23.5 which rules out this hypothesis at 99.9997 \% C.L. The reason for
the difference between the C.L. at which oscillations/transitions into
active and into sterile neutrinos are ruled out is the absence in the
latter case of neutral current contribution to the neutrino-electron
scattering rate which exacerbates the difficulty in describing
simultaneously the Homestake and Kamiokande data.

The constraints given above depend on the solar model used in the
analysis. Of the three recent solar models \cite{BP95,CDFLR,RVCD} only
Ref.\cite{BP95} gives a complete description of the estimated
uncertainties in the neutrino fluxes related to the inaccurately known
input parameters. In order to study this dependence on the solar
model, as explained in the previous section, we disregard the solar
model uncertainties of the most uncertain of the major solar neutrino
fluxes, $\Phi_B$ and $\Phi_{Be}$. For $f_{Be}$ we chose three
representative values, $f_{Be} = 0.7, 1.0$ and $1.3$. In each of these
three separate cases $f_B$ as well as $P$ are treated as free
parameters. For each of the three solar models \cite{BP95,CDFLR,RVCD}
we computed the $\chi^2$ (Eq. \ref{chi}) for a large number of pairs
$(f_B, P)$ from the following intervals: $0 \leq f_{B} \leq 2$ and $0
\leq P \leq 1$, and found the minimum, $\chi^2_{min}$, for each of the
three representative values of $f_{Be}$. The value of $\chi^2_{min}$
determine the C.L. at which the tested hypotheses are ruled out by the
data. The results are summarized in Table \ref{chires} for the solar
models \cite{BP95,CDFLR,RVCD} and for oscillations/transitions into
active as well as into sterile neutrinos.

For $f_{\rm Be} = 1$ the energy independent solar $\nu_e$ suppression
mechanisms are ruled out at (96.5)\% C.L. ($\chi^2 = 6.8$ (2 d.f.))
for all three solar models studied here.  Even if we relax the
constraint on the $^7{\rm Be}$ flux, the energy independent mechanisms
still give a poor fit to the data. The smallest minimum $\chi^2_{min}
= 5.9$ for oscillations/transitions into active neutrinos is achieved
for $f_{Be} = 0.7$.  This rules out any energy independent mechanism
as a solution of the solar neutrino problem at 94.5 \% C.L. (2
d.f.). Higher values of $f_{Be}$ are ruled out at even higher ($\simeq
97.5$ \% for $f_{Be} = 1.3$) C.L. The constraints on neutrino
oscillations/transitions into sterile neutrinos are somewhat stronger,
e.g., for $f_{Be} = 0.7$ the $\chi^2_{min}$ is 8.2. This rules out
such mechanisms at (at least) 98 \% C.L.

The week dependence of the results on the assumed beryllium neutrino
flux shows that the residual theoretical uncertainties, which we have
neglected here and which are related to all other neutrino fluxes
except $^8{\rm B}$ and $^7{\rm Be}$, cannot significantly change these
results. Those uncertainties are either considerably smaller ($pp$ and
$pep$ neutrinos) than the uncertainties in the $^{8}$B and $^{7}$Be
neutrino fluxes, which have been allowed to vary outside their
estimated $3\sigma$ limits, or are related to sources (CNO), which
give minor contributions to the event rates in the operating
experiments. In several test cases we have verified this by a complete
calculation where for the model \cite{BP95}, in addition to varying
$f_B$ and $P$, we took into account the residual uncertainties in the
CNO, $pep$ and $pp$ neutrino fluxes.

In our analysis we did not include solar models without diffusion,
e.g., \cite{turck}. These models are strongly disfavored by
helioseismological data \cite{helio,BPin} although the predicted
neutrino fluxes are only slightly lower than in the more advanced
models included in Table \ref{chires}. We also did not include results
for the model in ref. \cite{DS}. The results for this model, which
differs from the models \cite{BP95,CDFLR,RVCD} mainly by the choice of
the astrophysical factors, give somewhat lower $\chi^2_{min}$ in all
cases changing the C.L. by less than 0.5$\sigma$.

The constraints so obtained on the energy independent solar $\nu_e$
suppression mechanisms depend only weakly on the particular solar
model used as a reference.  In addition, we want to emphasize that at
present there is no compelling reason to expect such strong
deviations, as the ones considered here, from the predicted fluxes in
standard solar models. Our goal was to show that even ad hoc changes
in the solar neutrino fluxes cannot avert our main result, namely that
the neutrino oscillation/conversion mechanisms which predict energy
independent suppression are strongly disfavored by the data from the
operating solar neutrino experiments.

In order to test the significance of the data from each solar neutrino
experiment for the poor fit provided by energy independent neutrino
oscillations/transition mechanisms we have computed the $\chi^2$ for
each of the three possible pairs of experiments, namely (ClAr + $(\nu
e)$), ($(\nu e)$ + GaGe) and (GaGe + ClAr). When only $P$ is allowed
to vary and the neutrino fluxes are assumed to be the ones from the
SSM in Ref.\cite{BP95}, the minimum $\chi^2$ for the three cases,
assuming oscillations/transitions into active neutrinos, are:

\begin{mathletters}
\begin{eqnarray}
{\rm ClAr} + (\nu e) : & \chi^2_{min} = 0.07; & ~~P = 0.28, \\ \label{ClK}
(\nu e) + {\rm GaGe} : & \chi^2_{min} = 2.22; & ~~P = 0.50, \\ \label{KGa}
{\rm GaGe} + {\rm ClAr}: & \chi^2_{min} = 10.7; & ~~P = 0.46. \label{GaCl}
\end{eqnarray}
\end{mathletters}
\noindent For oscillations/transitions into sterile neutrinos we obtain:

\begin{mathletters}
\begin{eqnarray}
{\rm ClAr} + (\nu e) : & \chi^2_{min} = 1.15; & ~~P = 0.34, \\ \label{ClKst}
(\nu e) + {\rm GaGe} : & \chi^2_{min} = 0.79; & ~~P = 0.50, \\ \label{KGast}
{\rm GaGe} + {\rm ClAr}: & \chi^2_{min} = 10.7; & ~~P = 0.46. \label{GaClst}
\end{eqnarray}
\end{mathletters}
{}From these results it is evident that (both for
oscillations/transitions into active and sterile neutrinos) when only
(GaGe + ClAr) experiments are considered the energy independent
solutions are strongly disfavored by the data whereas both (ClAr +
$(\nu e)$) and (GaGe + ($\nu e$)) data can be fitted by an energy
independent suppression.

When both $P$ and the boron neutrino flux are allowed to vary for any
pair of experimental results there is always an acceptably small
$\chi^2_{min}$. The values of $P$ and $f_B$ for which $\chi^2$ is
minimum are slightly different for the different solar
models. However, the assumption that both the SSM and at least one of
the solar neutrino experiments are wrong, seems too extreme and hardly
justifiable.

Some energy independent mechanisms predict a fixed value of the
survival probability which cannot be varied. Examples of such
mechanisms are neutrino oscillations with maximal mixing between two
(P = 1/2) or three (P = 1/3) neutrinos \cite{P2}, as well as the
mechanism discussed in ref. \cite{harrison} (P = 5/9). The results in
Table \ref{TableP} show that all three values of the survival
probability, P = 1/3, 1/2, 5/9, are ruled out by the data both for
oscillations/transitions into active and into sterile neutrinos.  The
$\chi^2_{min}$ values found correspond respectively to 99.4, 96.5,
94.5 \% C.L. and to 99.96, 99.4, 98.4 \% C.L. for the two types of
oscillations/transitions.  Not only are the minimum $\chi^2$ too high
for all three models considered, but the values of the boron and
especially of the beryllium neutrino fluxes, for which the C.L. is
lowest, are far outside the acceptable uncertainty ranges for these
fluxes.

\section{Conclusions}
\label{conclusions} Mechanisms proposed as solutions of the
solar neutrino problem and predicting an energy independent
suppression of the solar neutrino spectrum as a result of neutrino
oscillations/transitions into active (sterile) neutrinos or
antineutrinos are ruled out by the data from the four operating
experiments \cite{CHLOR,KAM,GALLEX,SAGE} at 99.96 (99.9997) \% C.L.,
if the standard solar model \cite{BP95} correctly predicts (within the
estimated uncertainties) the solar neutrino fluxes from the different
sources. Even relaxing the constraints on the $^{7}$Be neutrino flux
by assuming that it differs by 30 \% from it's values in three recent
solar models with heavy element diffusion, and treating the $^{8}$B
neutrino flux as a free parameter, still provides a poor fit to the
data for any value of the energy independent suppression
factor. Suppression factors of 1/3, 1/2, and 5/9 \cite{P2,harrison}
are ruled out by the data at 99.4, 96.5, 94.5 \% C.L.  (99.96, 99.4,
98.4 \% C.L.)  for oscillations/transitions into active (sterile)
neutrinos even if one treats both the $^{8}$B and $^{7}$Be fluxes as
free parameters in the analysis. Not only are the minimum $\chi^2$ too
high, but the lowest $\chi^2_{min}$ values are reached for boron and
especially beryllium neutrino fluxes which lie far outside the
acceptable uncertainty ranges for these fluxes.

\section{Acknowledgments} P.K. is thankfull to K. Babu and
J.N. Bahcall for very usefull and stimulating discussions.  The work
of P.K. was partially supported by NSF grant \#PHY-9513835, while the
work of S.T.P. was supported in part by the EEC grant ERBCHRX CT930132
and by grant PH-510 from the Bulgarian Science Foundation.

\newpage
\begin{table}
\caption{Solar neutrino data used in the analysis and the theoretical
predictions from the SSM \protect\cite{BP95}.  The units are SNU for
the event rates in the radiochemical detectors and ${\rm cm}^{-2}{\rm
s}^{-1}$ for the flux measured by Kamiokande. For both the chlorine
and gallium experiments the theoretically predicted event rates have
been updated following the results in Ref.\protect\cite{BLABFN}.}
\begin{tabular}{l c c c}
Experiment & Data                  & Theory  & Units\\ \hline
Chlorine    & $2.56 \pm 0.16 \pm 0.14$  & $9.5 {}^{+1.2}_{-1.4}$ & SNU \\
Kamiokande  & $2.80 \pm 0.19 \pm 0.33$  & $6.62 {}^{+0.93}_{-1.12}$
& $10^6~{\rm cm}^{-2}{\rm s}^{-1}$    \\
GALLEX     & $69.7 \pm 6.7 {}^{+3.9}_{-4.5}$    & $136.8 {}^{+8}_{-7}$
& SNU    \\
SAGE       & $72.0 {}^{+12}_{-10} {}^{+5}_{-7}$ & $136.8 {}^{+8}_{-7}$
& SNU    \\ \hline
\end{tabular}
\label{SNData}
\end{table}

\begin{table}
\caption{Minimum $\chi^2$ for the energy independent mechanisms
including oscillations/transitions into active or sterile
neutrinos. The results are given for three reference solar models
\protect\cite{BP95,CDFLR,RVCD}. The $^7{\rm Be}$ neutrino flux has
been chosen: a) equal to the one in the corresponding solar model, b)
30 \% higher and c) 30 \% lower. The values of $f_B$, the ratio of the
boron neutrino flux to the predicted $^8{\rm B}$ flux in the
relevant SSM and $P$, the energy independent suppression factor,
for which $\chi^2_{min}$ is achieved in each case, are given also.}
\begin{tabular}{l c c c c c c c}
 & & \multicolumn{3}{c}{active} & \multicolumn{3}{c}{sterile} \\
\cline{3-8}
solar model & $f_{Be}$ & $\chi^2_{min}$ & $f_B$ & $P$ &
$\chi^2_{min}$ & $f_B$ & $P$  \\
\hline
            & 0.7      & 6.6 & 0.47 & 0.55  & 9.2  & 0.47 & 0.56 \\
BP'95       & 1.0      & 7.5 & 0.51 & 0.49  & 10.7 & 0.49 & 0.51 \\
            & 1.3      & 8.2 & 0.54 & 0.45  & 12.0 & 0.51 & 0.47 \\
\hline
            & 0.7      & 5.9 & 0.61 & 0.57  & 8.2  & 0.60  & 0.58 \\
CDFLR       & 1.0      & 6.8 & 0.64 & 0.52  & 9.6  & 0.62  & 0.54 \\
            & 1.3      & 7.5 & 0.67 & 0.48  & 10.8 & 0.64  & 0.50 \\
\hline
            & 0.7      & 6.7 & 0.50 & 0.56  & 9.3  & 0.49  & 0.57 \\
RVCD        & 1.0      & 7.6 & 0.54 & 0.50  & 10.9 & 0.52  & 0.52 \\
            & 1.3      & 8.4 & 0.57 & 0.46  & 12.3 & 0.54  & 0.48 \\
\end{tabular}
\label{chires}
\end{table}

\newpage
\begin{table}
\caption{Minimum $\chi^2$ and the corresponding values of $f_B$ and
$f_{Be}$ for three characteristic values of $P$ (the solar $\nu_e$
survival probability) in specific models predicting energy independent
suppression of the solar neutrino spectrum. The results are given for
three solar models \protect\cite{BP95,CDFLR,RVCD}. In each case the
upper row is for oscillations/transitions into active neutrinos,
whereas the lower row is for oscillations transitions into sterile
neutrinos.}
\begin{tabular}{l c c c c c c c c c}
 & \multicolumn{3}{c}{P = 1/3} & \multicolumn{3}{c}{P = 1/2} &
\multicolumn{3}{c}{P = 5/9} \\ \cline{2-10}
solar model & $\chi^2_{min}$ & $f_B$ &
$f_{Be}$ & $\chi^2_{min}$ & $f_B$ & $f_{Be}$ & $\chi^2_{min}$ & $f_B$
& $f_{Be}$ \\ \hline

BP'95 & 10.3 &  0.71 & 2.0  &  ~7.0 &  0.53 &  0.75 &  6.0  &   0.50 &  0.44 \\
      & 16.2 &  0.74 & 2.0  &  10.1 &  0.55 &  0.67 &  8.3  &   0.52 &  0.36 \\
\hline
CDFLR & 11.1 &  0.97 & 2.0  &  ~6.8 &  0.69 &  0.91 &  5.8  &   0.65 &  0.55 \\
      & 16.5 &  1.01 & 2.0  &  ~9.8 &  0.72 &  0.81 &  8.1  &   0.67 &  0.48 \\
\hline
RVCD  & 11.1 &  0.78 & 2.0  &  ~7.4 &  0.57 &  0.78 &  6.3  &   0.53 &  0.48 \\
      & 17.2 &  0.81 & 2.0  &  10.6 &  0.59 &  0.70 &  8.8  &   0.55 &  0.40  \\
\hline
\end{tabular}
\label{TableP}
\end{table}

\end{document}